\documentclass[runningheads]{llncs}
\usepackage{graphicx}
\usepackage{enumitem}
\usepackage{tabularx}
\usepackage{hyperref}
\usepackage{amssymb}
\usepackage{subcaption}
\usepackage{stmaryrd}
\usepackage{booktabs}
\usepackage{mathtools} 
\usepackage{tkz-euclide}
\usepackage{comment}
\usepackage{wrapfig}
\usepackage[bottom]{footmisc}
\usepackage[misc]{ifsym}
\usepackage{cite}
\usepackage{rotating}
\usetikzlibrary{arrows.meta}
\usetikzlibrary{calc}

\newcommand{\objects}{\mathcal{G}}

\newcommand{\attributes}{\mathcal{M}}
\newcommand{\incidence}{\mathcal{I}}
\newcommand{\formalcontext}{(\objects, \attributes, \incidence)}

\begin{document}

\title{Exploiting Formal Concept Analysis for Data Modeling in Data Lakes}

\author{Anes Bendimerad\inst{1}
\and
Romain Mathonat\inst{1}
\and
Youcef Remil\inst{1}
\and
Mehdi Kaytoue\inst{1}
}

\institute{Infologic R\&D, 26500 Bourg-Lès-Valence, France  
 \\ \email{mka@infologic.fr}}

\maketitle

\begin{abstract}
Data lakes are widely used to store extensive and heterogeneous datasets for advanced analytics. However, the unstructured nature of data in these repositories introduces complexities in exploiting them and extracting meaningful insights.  This motivates the need of exploring efficient approaches for consolidating data lakes and deriving a common and unified schema. This paper introduces a practical data visualization and analysis approach rooted in Formal Concept Analysis (FCA) to systematically clean, organize, and design data structures within a data lake. We explore diverse data structures stored in our data lake at Infologic, including InfluxDB measurements and Elasticsearch indexes, aiming to derive conventions for a more accessible data model. Leveraging FCA, we represent data structures as objects, analyze the concept lattice, and present two strategies—top-down and bottom-up—to unify these structures and establish a common schema. Our methodology yields significant results, enabling the identification of common concepts in the data structures, such as ``resources'' along with their underlying shared fields ($timestamp$, $type$, $usedRatio$, etc.). Moreover, the number of distinct data structure field names is reduced by 54\% (from 190 to 88) in the studied subset of our data lake. We achieve a complete coverage of 80\% of data structures with only 34 distinct field names, a significant improvement from the initial 121 field names that were needed to reach such coverage. The paper provides insights into the Infologic ecosystem, problem formulation, exploration strategies, and presents both qualitative and quantitative results. The source code and datasets of this work are made available: \url{https://zenodo.org/records/10589722}

\end{abstract}

\section{Introduction}\label{sec:introduction}

Organizations increasingly rely on data lakes~\cite{DBLP:journals/jiis/SawadogoD21} as versatile repositories to store vast and heterogeneous datasets for advanced analytics. The flexibility and scalability offered by data lakes have positioned them as a bedrock for managing massive volumes of raw, unstructured, and heterogeneous data. As defined in~\cite{DBLP:journals/pvldb/NargesianZMPA19}, a data lake is a massive collection of datasets that: (1) may be hosted in different storage systems; (2) may vary in their formats; (3) may not be accompanied by any useful metadata or may use different formats to describe their metadata; and (4) may change autonomously over time. At Infologic~\cite{infologiccopilote}, our data lake serves as a key component in the scope of our predictive maintenance system~\cite{DBLP:conf/sigsoft/BendimeradRMK23}, enabling seamless aggregation of continuously collected data from diverse sources at a low cost. New data collections can be easily added to the data lake by different teams, without the need of defining a priori the data schema. Nevertheless, the unstructured and heterogeneous nature of the data stored in data lakes poses a significant challenge, hindering the full exploitation of their inherent value~\cite{DBLP:journals/pvldb/NargesianZMPA19}. It becomes difficult to have a clear understanding of the content of the data and to implement data pipelines that generalize well. Furthermore, executing some common analytics operations between data structures, such as merges and joins, becomes cumbersome, as these structures may use different names for the same fields. Particularly at Infologic, we were using two storage systems, InfluxDB~\cite{influxdb} and Elasticsearch~\cite{elasticsearch}, each adhering to distinct conventions. This combination of two storage systems were motivated by the effectiveness of InfluxDB in handling metrics and time series, while Elasticsearch is more efficient as a search engine, especially in textual data and JSON documents. 

We aim to explore the data structures within our data lake, and extract a set of conventions to consolidate our data model. The data structures under examination include the schemas of InfluxDB measurements and Elasticsearch indexes.
Deriving a common schema is a problem that has interested many practitioners. Notably, the Elasticsearch community has proposed the ECS~\cite{ecs} (Elastic Common Schema) to define a common set of fields to be used when storing events data in Elasticsearch. The Common Event Format~\cite{arcsight2009common} (CEF) has been designed to propose standard naming conventions for logs in network and security devices and computer systems. 
Inspired by these well-established standards, we seek to derive a tailored data model that not only aligns with industry practices, but also accommodates the specificities of our business at Infologic. Such specificities concern the architecture that governs our Copilote ERP software, as well as conventions that are already followed in the Relational database that is used to store Copilote critical business data.

This paper addresses this challenge through a comprehensive exploration of a novel approach grounded in Formal Concept Analysis (FCA)~\cite{wille1982restructuring,GanterW99}, aimed at systematically cleaning, structuring, and designing the data within our data lake. We perform interactive data analysis, leveraging the concept lattice as a central tool. FCA has been exploited to address various challenges in both software engineering and data engineering, such as mining functional dependencies for SQL data refractoring~\cite{DBLP:journals/amai/BaixeriesKN14,DBLP:journals/jetai/LopesPL02}, creating and merging of ontology top-levels~\cite{DBLP:conf/iccs/GanterS03}, and fault localization in software~\cite{DBLP:conf/icfca/CellierDFR08}.  However, our paper represents the first attempt to employ the concept lattice as a visual tool for consolidating structures in data lakes. We represent data structures, including tables, measurements, and indexes, as objects within a formal context. Each of these objects is described by Boolean attributes that indicate whether a field is present in the related data structure. Subsequently, we derive and analyze the concept lattice from this formal context. In our exploration of this lattice, we present two distinct strategies—top-down and bottom-up—that leverage visual insights to unify field names and establish a common schema. Our methodology yields significant results, enabling the identification and unification of common concepts in the data structures, such as ``resources'' along with their underlying shared fields ($timestamp$, $type$, $usedRatio$, etc.).
Moreover, the number of data structure field names is reduced by 54\%, from 190 to 88 in a subset of our data lake. We achieve a complete coverage of 80\% of data structures in this subset with only 34 distinct field names, a significant improvement from the initial 121 field names that were needed to reach such coverage.

\noindent \textbf{Outline.} Section~\ref{sec:background} provides an overview of Infologic and its ERP software, Copilote, accompanied by a description of our data lake that is used to store predictive maintenance data. Section~\ref{sec:problem} formulates the studied dataset and problem within the FCA framework, and describes the generation of the concept lattice by illustrating the process with a toy example. In Section~\ref{sec:exploting_lattice}, we show the used strategies to explore the concept lattice and derive insights that guide us in building our data model. Section~\ref{sec:results} approaches the final results from a qualitative and quantitative points of view. Section~\ref{sec:conclusion} provides a conclusion and future avenues.
\begin{figure}[t]
	\centering
	\includegraphics[width=1\linewidth]{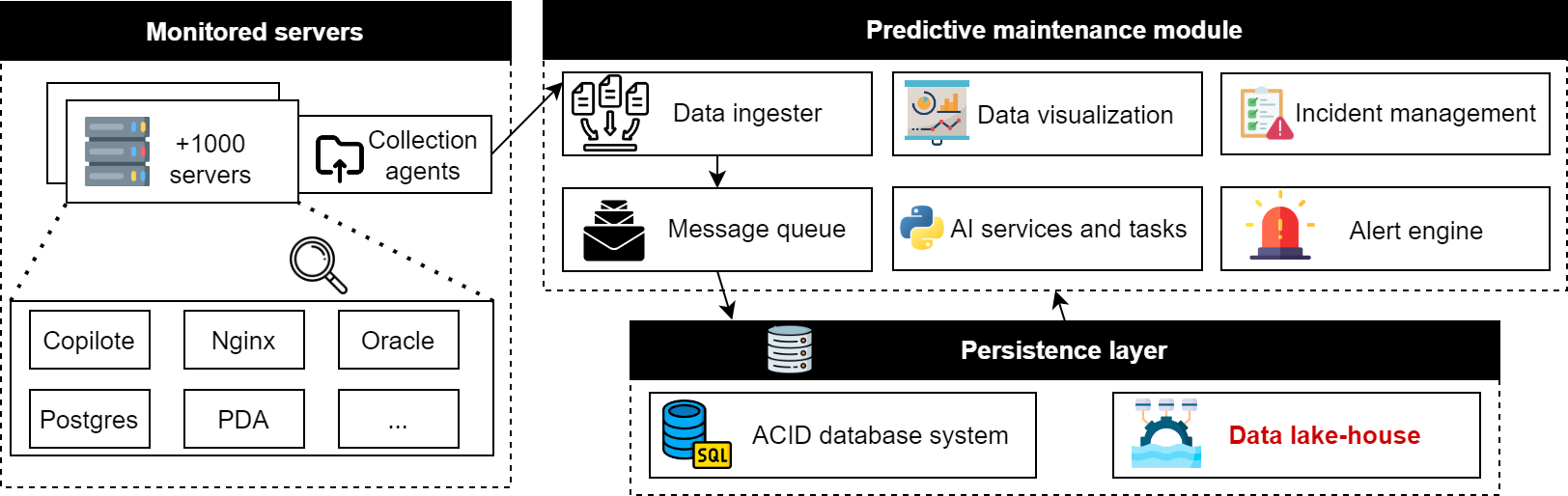}
	\caption[Simplified architecture of predictive maintenance at Infologic.]{Simplified architecture of predictive maintenance at Infologic~\cite{DBLP:conf/sigsoft/BendimeradRMK23}.}%
	\label{fig:schema_aiops}
\end{figure}

\section{Background}\label{sec:background}

\subsection{Infologic}\label{subsec:infologic}

Infologic~\cite{infologiccopilote}  is a leading provider of enterprise resource planning (ERP) solutions for the agri-food, health nutrition, and cosmetic sectors in France. Its flagship product, Copilote, is an ERP software designed to optimize and automate a large panel of business processes, including sales tracking, supply chain and customer relationship management. {\sc Infologic} provides maintenance of the ERP instances and infrastructure in operation for its clients. As the proper functioning of their businesses depends heavily on the reliable performance and accessibility of the ERP, it is crucial to ensure high availability and excellent maintenance for Copilote. To this aim, Infologic has made substantial investments in a predictive maintenance project~\cite{DBLP:conf/sigsoft/BendimeradRMK23,DBLP:conf/kbse/RemilBMCK21, DBLP:conf/dsaa/RemilBPRK21, remil2023mining, DBLP:journals/corr/abs-2310-06703, remil2023data}. In~\cite{DBLP:conf/sigsoft/BendimeradRMK23}, the architecture of this project has been presented, and its key components have been described in details. Figure~\ref{fig:schema_aiops} provides a simplified overview of this architecture. One of the foundational components of this project is the data lake, or its broader iteration, the data lake-house~\cite{DBLP:conf/cidr/Zaharia0XA21,harby2022data}, incorporating data warehousing features~\cite{DBLP:journals/sigmod/Anisimov03,DBLP:journals/dr/Widom99c}. 

\begin{figure}[t]
	\centering
	\includegraphics[width=0.85\linewidth]{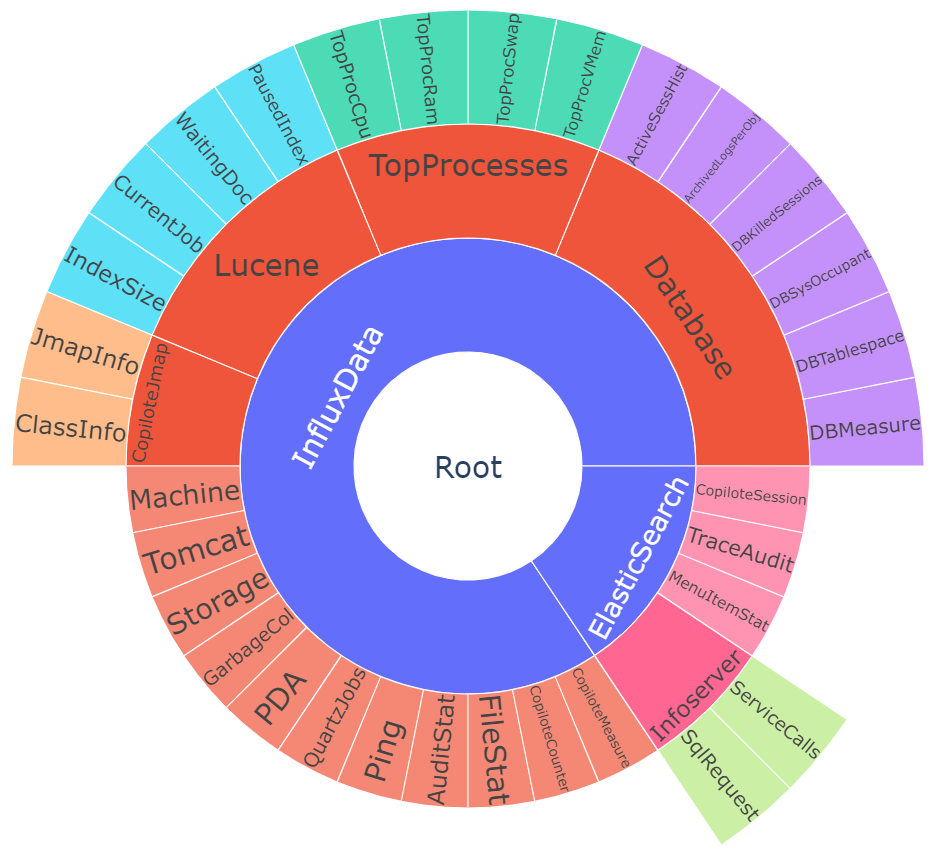}
	\caption[Data structures stored in our data lake and studied in this paper.]{Data structures stored in our data lake and studied in this paper.}%
	\label{fig:datalake}
\end{figure}

\subsection{Data lake} \label{subsec:datalake}
During the period of our study, the primary components of our data lake comprised InfluxDB~\cite{influxdb} and Elasticsearch~\cite{elasticsearch}, serving as repositories for the continuous collection and storage of diverse datasets. In Figure~\ref{fig:datalake}, we present a subset of data structures contained in our data lake, which constitutes the focus of our analysis. This figure depicts our data as a hierarchy whose leaves represent specific data structures, such as $Machine$, $Tomcat$, $Storage$, among others. The internal nodes in the figure denote cohesive groups of structures belonging to distinct domains. For instance, $Lucene$ forms a group encompassing four InfluxDB tables (measurements) designed for storing Lucene monitoring data~\cite{lucene2010apache}, including $PausedIndex$, $WaitingDoc$, $CurrentJob$, and $IndexSize$. Time series data were stored in InfluxDB, whereas Elasticsearch was dedicated to textual data and JSON documents.
The examined  subset comprises 32 data structures (InfluxDB measurements and Elasticsearch indexes) utilizing 190 distinct field names. Some of these names are used in several data structures. 
\begin{table}[t]

\centering
\caption{Toy example of data structures with their underlying fields\label{tab:formalContext}.}
  \begin{tabular}{c||c|c|c|c|c|c|c|c}
  \hline   
  $\:\objects\:$ (structures)  & time & timestamp  & used & max & path & name & serviceName & duration \\ 
   \hline
   Storage & $\times$ &  & $\times$ & $\times$ & $\times$ & & &  \\ 
   DBTablespace & $\times$ & & $\times$ & $\times$ & & $\times$ & &  \\ 
   ServiceCall &  & $\times$ & &  & & $\times$ & $\times$  & $\times$   \\ 
   \hline
 \end{tabular}
\end{table}

\section{Problem Formulation}\label{sec:problem}

\subsection{Data model with FCA}\label{subsec:dataset}

We formalize the dataset describing our data structures based on Formal Concept Analysis~(FCA)~\cite{wille1982restructuring,GanterW99}.
We define the formal context~\cite{GanterW99} as a triple $\mathbb{K} = \formalcontext$  comprising two sets $\objects$ and $\attributes$ and an incidence relation $\incidence$ between $\objects$ and $\mathcal{M}$. Elements of $\objects$ are called objects, and elements of $\mathcal{M}$ are called attributes. To signify that an object  $g \in \objects$ has an attribute $m \in \mathcal{M}$, we use the notation $g \incidence m$. Table~\ref{tab:formalContext} reports a formal context $\formalcontext$ where objects in $\objects$ represent data structures from the data lake, and attributes in $\mathcal{M}$ denote fields within these data structures. The incidence relation $\incidence$ is visually depicted by crosses in the table, and it represents the fact that a data structure contains a field. For instance, we have ``$Storage$ $\incidence$ $used$'' that can be read as: ``the $Storage$ data structure contains the field $used$". In total, the data structure $Storage$ is characterized by the following fields: $time, used, max, path$. Notably, some fields in data structures of Table~\ref{tab:formalContext} convey similar meanings but are designated by distinct names, such as $time$ and $timestamp$, or $serviceName$ and $name$. The goal of our study is to analyze a comprehensive set of data structures, and identify groups of akin field names that manifest recurrently and signify the same underlying notion. Subsequently, we aim to generalize these field names uniformly, establishing a cohesive and unified schema.

Two fundamental operators, namely extent and intent, are defined on a formal context $\mathbb{K} = \formalcontext$. The extent operator, denoted $ext$, associates to each subset of attributes $B \subseteq \mathcal{M}$ the set of objects $g \in \objects$ possessing all attributes in $B$, that is, $ext(B)=\{g \in \objects \mid (\forall m \in B)\: g\,\mathcal{I}\,m\}$. Dually, the intent operator, denoted $int$, associates to each subset of objects $A \subseteq \mathcal{\objects}$ the set of attributes $m \in \mathcal{M}$ shared among the objects in $A$, that is, $
int(A) = {\{m \in \mathcal{M} \mid (\forall g \in A)\: g\,\mathcal{I}\,m\}}$. It is noteworthy that, for $B \subseteq \attributes$ and $A \subseteq \objects$,  the following relationships hold:
$ext(B) = \bigcap_{m \in B} ext(\{m\})$  and  
$int(A) = \bigcap_{g \in A} int(\{g\})$.
A key theorem in FCA (Proposition~10 in \cite{GanterW99}) is:
\begin{theorem}\label{thm:galoisConnection}
The pair of functions $(ext, int)$ form a Galois connection between the power set lattices $(2^{\objects}, \subseteq)$ and $(2^{\mathcal{M}}, \subseteq)$. That is, $ext \circ int$ and $int \circ ext$ are closure operators on $(2^{\objects}, \subseteq)$ and, $(2^{\mathcal{M}}, \subseteq)$ respectively. 
\end{theorem}

Following this theorem, we can build a concept lattice $(\mathfrak{B}(\mathbb{K}), \leq)$. Elements of $\mathfrak{B}(\mathbb{K})$ are formal concepts and are of the form $(A, B) \in 2^{\objects} \times 2^{\attributes}$ with $A = ext(B)$ and $B = int(A)$. 
In Table~\ref{tab:formalContext}, $ext(\{time, used, max\}) = \{Storage, DBTablespace\}$, meaning that objects possessing the fields $time$, $used$, and $max$ are $Storage$ and $DBTablespace$. Dually, $int(\{Storage, DBTablespace\}) = \{time, used, max\}$, indicating that the common fields between $Storage$ and $DBTablespace$ are $time$, $used$, and $max$. Since $int(ext(\{time, used, max\})) = \{time, used, max\}$, the pair $(\{Storage, DBTablespace\}$, $\{time, used, max\})$ is a formal concept.

%

\begin{figure}[t]
  \centering
  \begin{subfigure}{0.4\textwidth}
    \includegraphics[width=\linewidth]{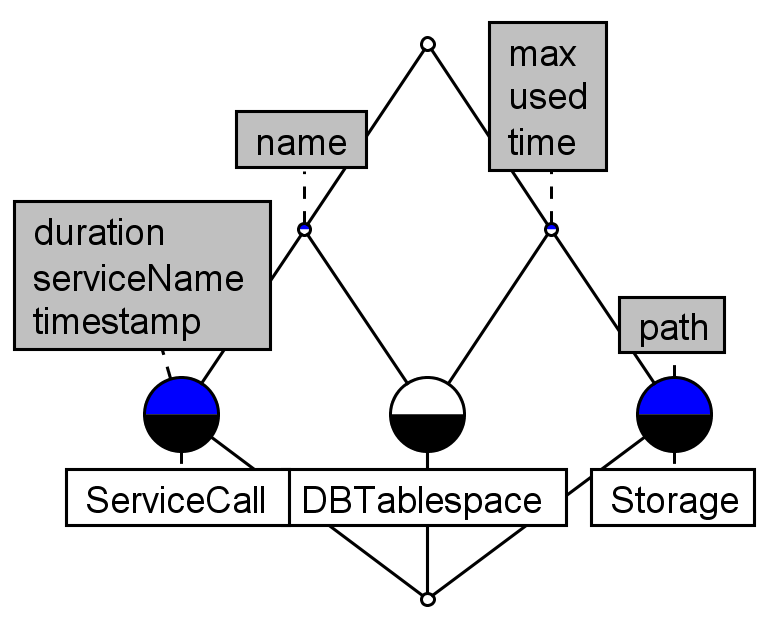}
    \caption{Before consolidating field names.}
  \end{subfigure}
  \hfill
  \begin{subfigure}{0.4\textwidth}
    \includegraphics[width=\linewidth]{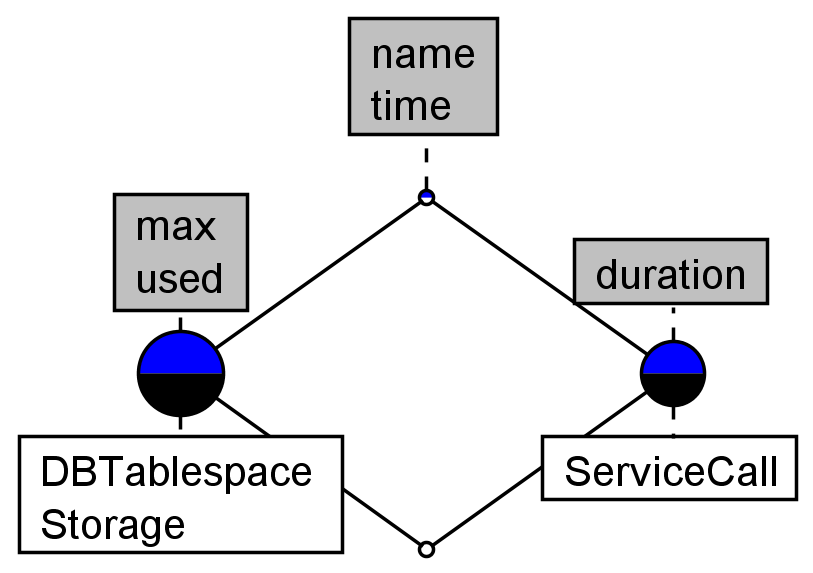}
    \caption{After consolidating field names.}
  \end{subfigure}
  \caption{The concept lattice before and after unifying fields from Table~\ref{tab:formalContext}.}
  \label{fig:toyLattice}
\end{figure}

\subsection{Concept lattice}\label{subsec:concept_lattice}
We construct the concept lattice $(\mathfrak{B}(\mathbb{K}), \leq)$ from our formal context $\mathbb{K}$. Various tools can be used to visualize the concept lattice given any formal context stored in some specific format~\cite{DBLP:conf/cla/SaabH022}, such as a CSV file. In our study, Concept Explorer~\cite{conexp} was utilized for this purpose. Figure~\ref{fig:toyLattice}~(a) shows the concept lattice generated from the toy dataset in Table~\ref{tab:formalContext}. Each node of this lattice represents a formal concept $(A, B) \in 2^{\objects} \times 2^{\attributes}$ such that $A = ext(B)$ and $B = int(A)$. For example, the right child of the root corresponds to the formal concept $(\{Storage, DBTablespace\},\{time, used, max\})$. Subsequently, the right child of the latter formal concept is $(\{Storage\},\{time, used, max, path\})$, which is a formal concept covering only the object (data structure) $Storage$. Using such data visualization, our aim is to analyze concepts in order to derive relevant unification of fields and structures. In the toy dataset, we can unify the field names $time$ and $timestamp$, renaming both as $time$. We can also consolidate the field names $serviceName$, $name$, and $path$ into a unified label, such as $name$. These transformations result in the unified field names ascending in the direction of the top of the lattice. Figure~\ref{fig:toyLattice}~(b) illustrates the final lattice after applying these transformations on the toy formal context of Table~\ref{tab:formalContext}. The two fields $name$ and $time$ have ascended to the root of the lattice since they are covered by all the objects $Storage$, $DBTablespace$, and $ServiceCall$. An alternative method to consolidate the data model is to exploit association rules derived from the formal context. However, we believe that the visual exploitation of the lattice is easier for FCA non-practitioners. We show in Section~\ref{sec:exploting_lattice} that extracting relevant insights is simplified when viewing the concept lattice.

\begin{figure}[t]
	\centering
	\includegraphics[width=1\linewidth]{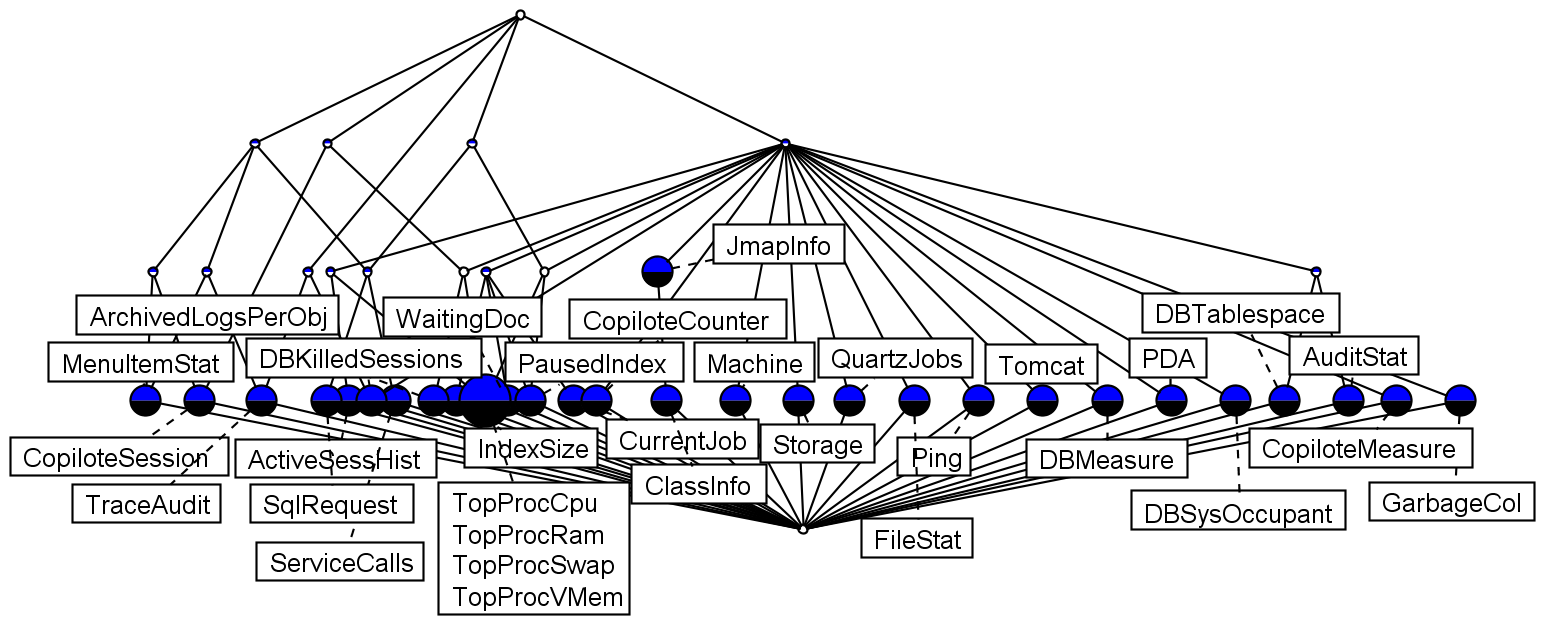}
	\caption[Concept lattice depicting data structures within our data lake. The objects (data structures) are indicated in the lattice.]{Concept lattice depicting data structures within our data lake. The objects (data structures) are indicated in the lattice.}%
	\label{fig:latticePriorObjects}
\end{figure}

\begin{figure}[t]
	\centering
	\includegraphics[width=1\linewidth]{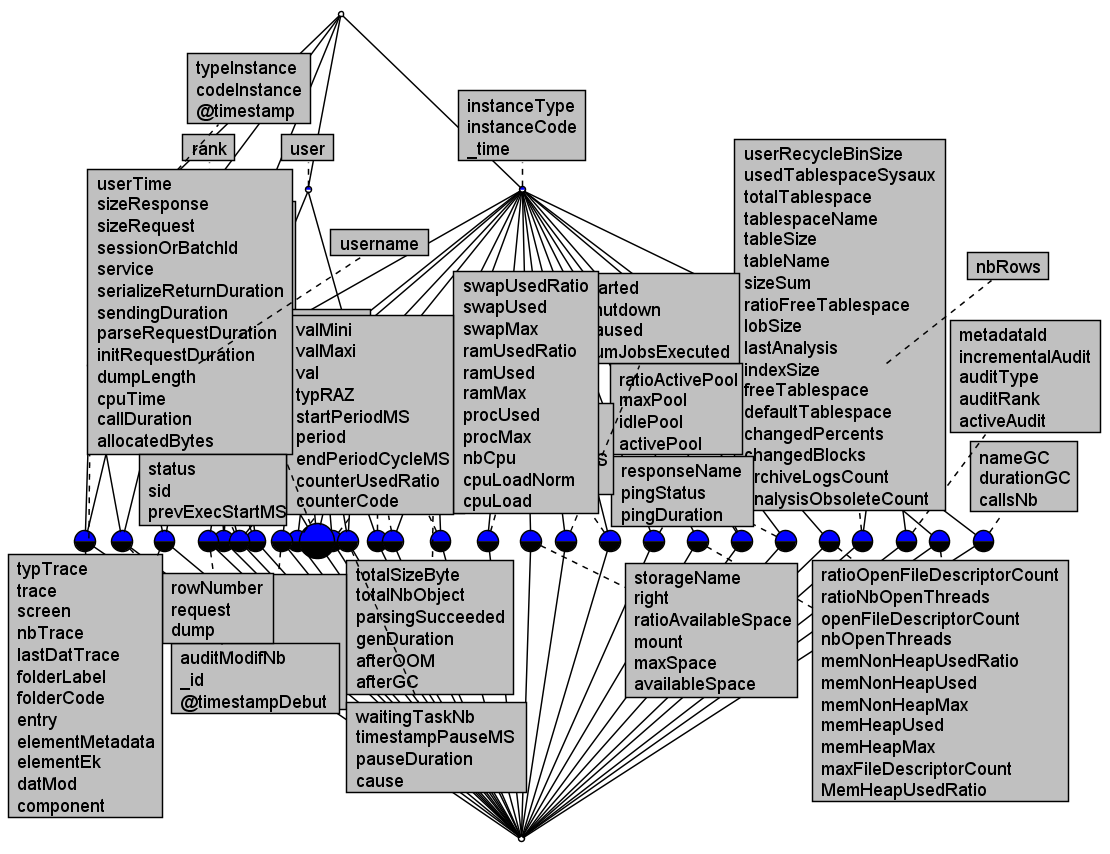}
	\caption[Concept lattice depicting data structures within our data lake. The objects are indicated in the lattice.]{Concept lattice depicting data structures within our data lake. The attributes (field names) are indicated in the lattice.}%
	\label{fig:latticePriorAttributes}
\end{figure}

\section{Exploiting the Concept Lattice}\label{sec:exploting_lattice}

\subsection{Analyzing the initial lattice}\label{sec:initial_lattice}

We generate the concept lattice from the dataset depicted in Figure~\ref{fig:datalake}, comprising 32 objects (data structures) and 191 attributes (field names). Figure~\ref{fig:latticePriorObjects} presents the complete lattice, highlighting the objects associated with each formal concept, while Figure~\ref{fig:latticePriorAttributes} displays the attributes within each concept. The root represents the formal concept that covers all the objects of the dataset. Since there is no field name that is present in all the data structures, the root node in Figure~\ref{fig:latticePriorAttributes} has an empty intent. Below the root node, we observe a few nodes that sit in this second level of the lattice, such as the concept with the intent $\{instanceType, instanceCode, \_time \}$. This concept reflects a set of field names that are present in every data structure of InfluxDB, but they are not used in Elasticsearch. Other examples of concepts as children of the root include those with the intent $\{rank\}$ and $\{user\}$. In Figure~\ref{fig:latticePriorObjects}, we notice that nearly all objects (data structures) reside in the penultimate level. Notably, this lattice  exhibits a small height and a large width due to the limited number of common field names between data structures, attributed to the absence of a standardized naming convention. Consequently, there is a limited number of internal formal concepts representing shared attributes. This lattice serves as the basis for our analysis, where we explore two distinct data analysis approaches outlined in this section: the top-down approach and the bottom-up approach.

\subsection{The top-down approach}\label{sec:top_down}
This approach consists in starting from the root (the top-node) of the hierarchy and find akin field names that can be unified. In this section, we illustrate this top-down approach with two interesting findings that were made possible thanks to this data exploration.

\begin{figure}[t]
  \centering
  \begin{subfigure}{0.4\textwidth}
    \includegraphics[width=\linewidth]{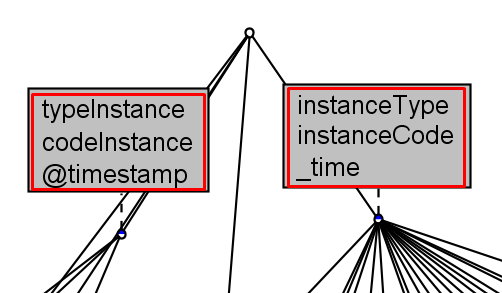}
    \caption{Before unifying generic fields.}
  \end{subfigure}
  \hfill
  \begin{subfigure}{0.4\textwidth}
    \includegraphics[width=\linewidth]{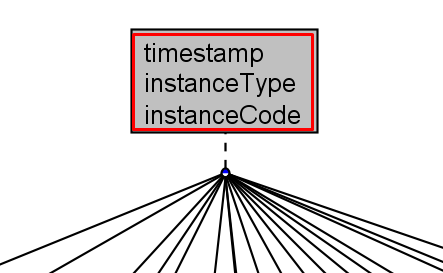}
    \caption{After unifying generic fields.}
  \end{subfigure}
  \caption{Top-down approach. Unifying generic fields from InfluxDB and Elasticsearch.}
  \label{fig:topdown1}
\end{figure}

\begin{figure}[t]
  \centering
  \begin{subfigure}{0.4\textwidth}
    \includegraphics[width=\linewidth]{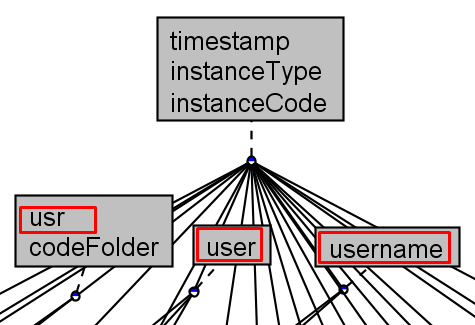}
    \caption{Before unifying user fields.}
  \end{subfigure}
  \hfill
  \begin{subfigure}{0.4\textwidth}
    \includegraphics[width=\linewidth]{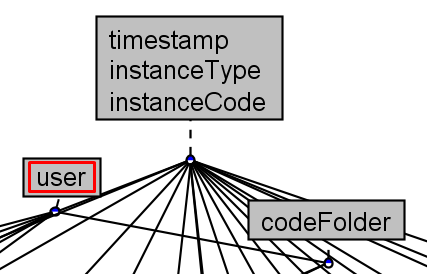}
    \caption{After unifying user fields.}
  \end{subfigure}
  \caption{Top-down approach. Unifying $user$ fields.}
  \label{fig:topdown2}
\end{figure}

\subsubsection{Unifying generic fields from InfluxDB and Elasticsearch.} 
As depicted in Figure~\ref{fig:topdown1}~(a), the concept lattice visualization provides a clear view of the root having two children covered by similar sets of fields. The first group, comprising $\{typeInstance, codeInstance, @timestamp\}$, is uniformly present in all data structures stored in Elasticsearch. Here, $typeInstance$ signifies the Copilote ERP instance type (production, test, development, deployment), $instanceCode$ is a unique identifier for Copilote instances, and $@timestamp$ denotes the creation time of the collected event or metric. Simultaneously, the second group $\{instanceType, instanceCode, \_time\}$ is utilized across all InfluxDB measurements. These two groups have been consolidated into the unified field names $\{timestamp, instanceType, instanceCode\}$. As shown in Figure~\ref{fig:topdown1}~(b), this group has ascended to the root of the lattice, since it is covered by all our data structures (all the objects of the formal context).

\subsubsection{Unifying the $user$ field.} Another compelling application of the top-down approach involves the consolidation of the field describing the user identifier, as illustrated in Figure~\ref{fig:topdown2}~(a). Initially, we observed distinct field names referencing the user identifier, namely $usr$, $user$, and $username$. Each of them is covered by several data structures, which makes them visible in the upper layers of the lattice. We have unified these identifiers under a standardized field name, $user$, as depicted in Figure~\ref{fig:topdown2}~(b).

\begin{figure}[t]
  \centering
  \begin{subfigure}{0.48\textwidth}
    \includegraphics[width=\linewidth]{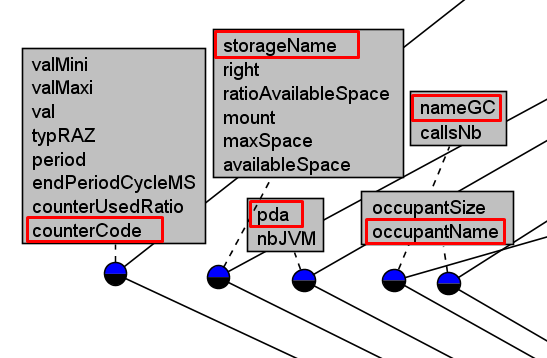}
    \caption{Before unifying code fields.}
  \end{subfigure}
  \hfill
  \begin{subfigure}{0.48\textwidth}
    \includegraphics[width=\linewidth]{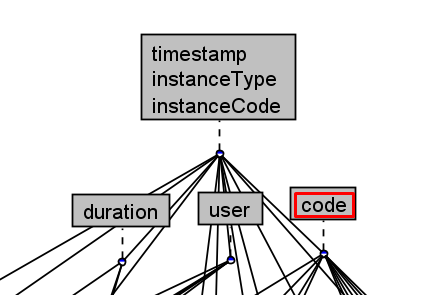}
    \caption{After unifying code fields.}
  \end{subfigure}
  \caption{Bottom-up approach. Unifying $code$ fields.}
  \label{fig:bottomup1}
\end{figure}

\subsection{The bottom-up approach}\label{sec:bottom_up}
Another method to exploit concept lattice visualization is to inspect the bottom layers to discern a substantial set of distinct fields that are used in a relatively limited set of data structures but represent the same notion. Unlike the top-down approach, which unifies groups of fields already used in many data structures, the bottom-up approach has another focus. It targets the unification of fields generally employed in unique data structures and, consequently, found only in the last layers of the lattice. We illustrate with three noteworthy  results that have been discovered through this methodology.

\subsubsection{Unifying the $code$ attribute.} 
By examining data structures in the last layer, we observe that many of them have a field dedicated to storing the name of a component monitored by the respective data structure.  Figure~\ref{fig:bottomup1}~(a) illustrates this observation with five data structures. For instance, $storageName$ denotes the name of the storage space monitored by metrics such as $maxSpace$ and $availableSpace$. Similarly, $nameGC$ represents the name of the Java Garbage Collector under surveillance, for which the number of executions ($callsNb$) is logged. We have unified these fields under  the same name, $code$, which subsequently ascended to a position beneath the root of the lattice, as shown in Figure~\ref{fig:bottomup1}~(b).

\begin{figure}[t]
  \centering
  \begin{subfigure}{0.48\textwidth}
    \includegraphics[width=\linewidth]{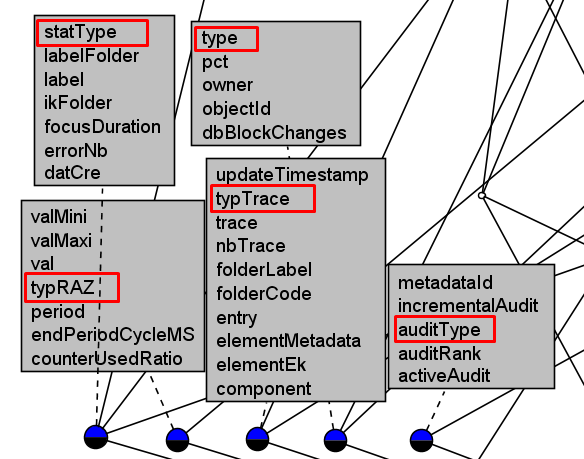}
    \caption{Before unifying type fields.}
  \end{subfigure}
  \hfill
  \begin{subfigure}{0.48\textwidth}
    \includegraphics[width=\linewidth]{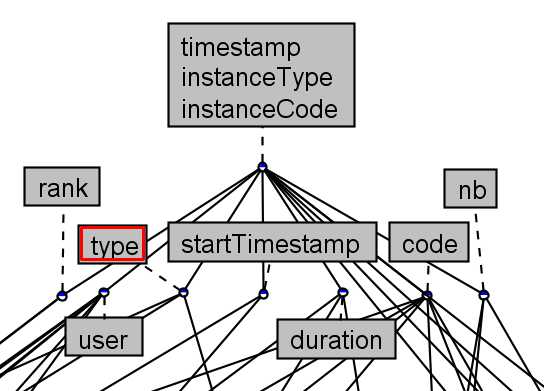}
    \caption{After unifying type fields.}
  \end{subfigure}
  \caption{Bottom-up approach. Unifying $type$ fields.}
  \label{fig:bottomup2}
\end{figure}

\begin{figure}[t]
  \centering
  \begin{subfigure}{0.48\textwidth}
    \includegraphics[width=\linewidth]{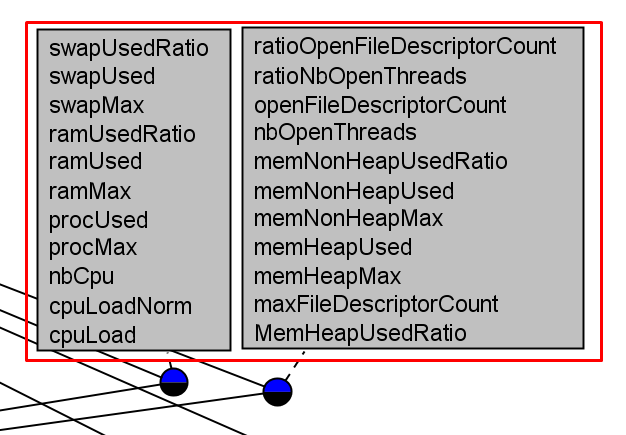}
    \caption{Before unifying resources fields.}
  \end{subfigure}
  \hfill
  \begin{subfigure}{0.4\textwidth}
    \includegraphics[width=\linewidth]{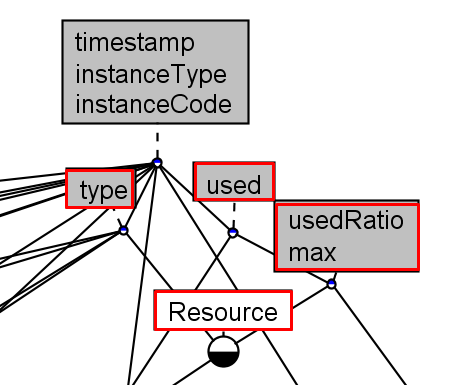}
    \caption{After unifying resources fields.}
  \end{subfigure}
  \caption{Bottom-up approach. Unifying resources fields in a $Resource$ table.}
  \label{fig:bottomup3}
\end{figure}

\subsubsection{Unifying the $type$ attribute.} 
An information that is present in a wide range of data structures is the type of the component described by the data. Figure~\ref{fig:bottomup2}~(a) showcases five selected examples of such fields: $statType$, $type$, $typRAZ$, $typTrace$, $auditType$. As reported in Figure~\ref{fig:bottomup2}~(b), all of these fields have been consolidated under the same name: $type$.

\subsubsection{Unifying the $Resource$ table.} 
In Figure~\ref{fig:bottomup3}, we report a specific optimization of data model. We show two data structures that describe resource utilization of the machine hosting the ERP, as well as its JVM. For example, $swapUsed$ indicates the used swap in bytes, $swapMax$ indicates the limit of the swap size, while $swapUsedRatio$ indicates the proportion of the used swap. Similar fields exist to describe other resources, such as RAM, Java heap space, Java non heap space, and more. We have consolidated these fields into a new data structure named $Resources$. Within this structure, the $type$ field indicates the resource being monitored (e.g., ``swap'', ``ram'', ``heap''). Subsequently, the utilization of the resource indicated in the $type$ field is described by three generic fields: $used$, $max$, and $usedRatio$. Figure~\ref{fig:bottomup3}~(b) reveals that all of these fields have ascended to the upper layers of the lattice, consequently elevating the entire $Resources$ structure within the lattice. 

\section{Results}\label{sec:results}
In this section, we begin by reporting and analyzing the final lattice derived from the dataset after performing our data structure consolidation. We highlight the distinctions between this final lattice and the one generated from the original dataset. Then, we perform a quantitative study to analyze the distribution of field names, their coverage of the data structures, and measure the improvement achieved from this perspective.

\begin{figure}[t]
	\centering
	\includegraphics[width=1\linewidth]{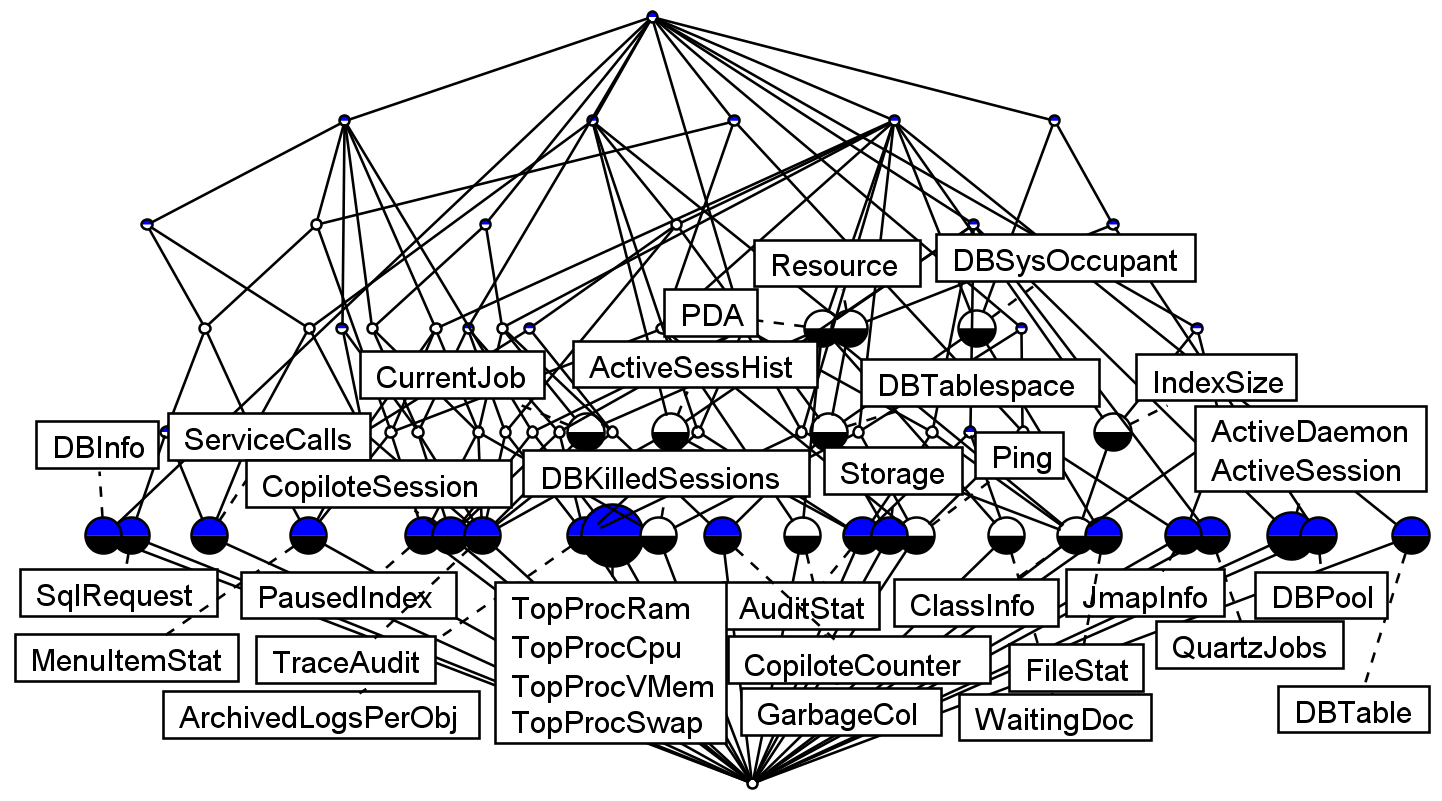}
	\caption{The concept lattice of our data structures after consolidating field names. The objects (data structures) are indicated in the lattice.}%
	\label{fig:latticeFinalObjects}
\end{figure}

\begin{figure}[t]
	\centering
	\includegraphics[width=1\linewidth]{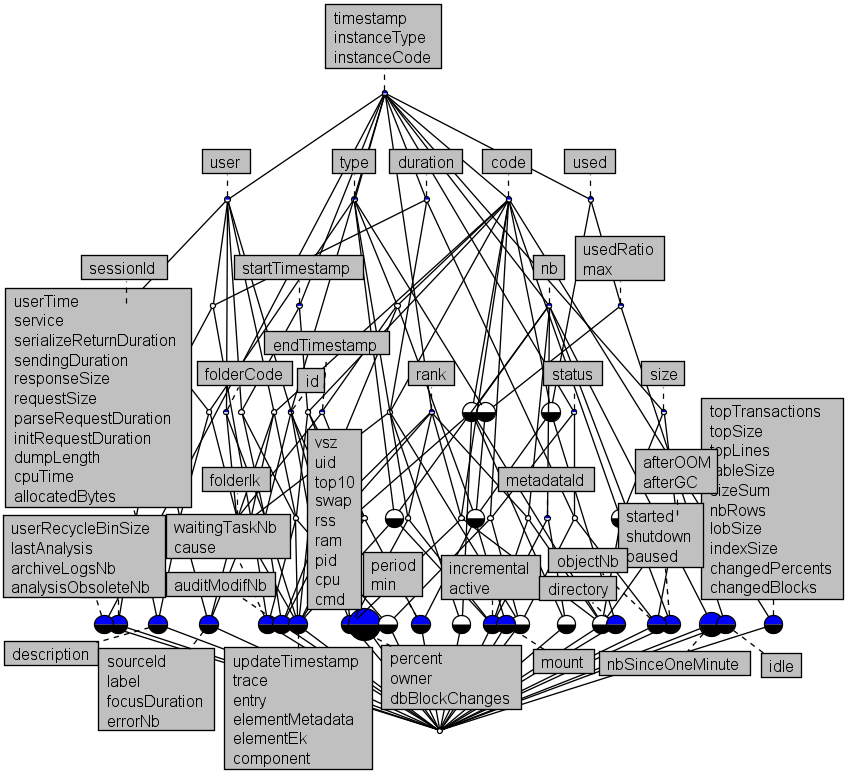}
	\caption{The concept lattice of our data structures after consolidating field names. The attributes (field names) are indicated in the lattice.}%
	\label{fig:latticeFinalAttributes}
\end{figure}

\subsection{Final lattice} \label{subsec:final_lattice}

Figure~\ref{fig:latticeFinalObjects} illustrates the concept lattice generated from the final dataset after applying all the transformations related to the proposed data model. The illustration also includes the objects associated to each of the formal concepts represented in the lattice. Some data structures have moved upward in the lattice, as they use only common fields without any specific field names. For example, $Resource$ emerges as one of the most generic data structures, positioned at one of the initial layers of the lattice. Figure~\ref{fig:latticeFinalAttributes} presents the same lattice but with a view on field names (attributes of the formal context). Three field names appear in the root of the lattice—$timestamp$, $instanceType$, and $instanceCode$—as they are covered by all the data structures. In the second layer, we find generic fields such as $user$, $type$, $duration$, $code$, $used$. Notably, the concept covered by $used$ represents resources whose consumption is measurable, and the field $used$ is employed to store the quantity of utilization of the related resource. For example, if the resource is the swap memory, $used$ indicates the number of bytes consumed by the swap. Furthermore, the concept covered by $used$ has a child that is extended with the fields $max$ and $usedRatio$, to represent resources whose capacity is limited and known. Another interesting observation in the lattice is that, while $duration$ sits in the second layer, the fields $startTimestamp$ and $endTimestamp$ are separated in other concepts that come lower in the lattice. A possible further consolidation of our model is to add the fields $startTimestamp$ and $endTimestamp$ to all the ``events'' data structures that are described by a $duration$. This would unify the three fields $duration$, $startTimestamp$ and $endTimestamp$ in a same formal concept. Following our data structure consolidation, the number of distinct field names has decreased from 190 to 88. However, the number of formal concepts has increased from 44 to 72, due to more sets of fields common between data structures. Consequently, the lattice height has increased from initially 4 to 6 after performing the data structure consolidation.

\begin{figure}[t]
	\centering
	\includegraphics[width=0.75\linewidth]{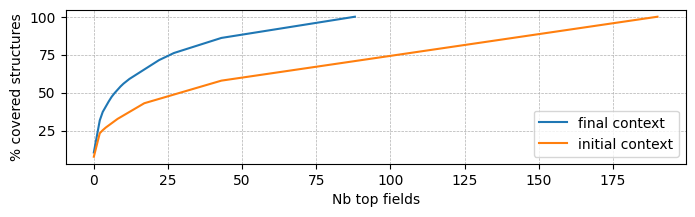}
	\caption[Covered data structures given the number of top field names used.]{Covered data structures given the number of top field names used.}%
	\label{fig:covered_relations}
\end{figure}

\subsection{Analyzing the number of attributes}
Our objective is to assess the extent of data structures coverage by the new fields names, and compare it with their coverage before applying our transformations. To achieve this, we sort the field names in descending order based on the frequency of their utilization in data structures. Subsequently, we measure the proportion of completely covered data structures by a given number of top fields ranked with respect to their utilization frequency. Figure~\ref{fig:covered_relations} reports the results. In the final context, we achieve a coverage of 75\% of data structures with 25 field names, marking a significant improvement compared to the initial dataset where the same number of field names covered less than 50\% of the data structures. Moreover, we can cover the entire dataset with 88 field names instead of original 190 field names, reducing them by 54\%.
\section{Conclusion}\label{sec:conclusion}

In conclusion, our application of Formal Concept Analysis (FCA) to the exploration of data lake structures at Infologic has proven highly effective. By systematically analyzing diverse data structures and leveraging FCA's concept lattice, we successfully reduced the number of distinct attributes by 54\%, from 190 to 88, and covered 80\% of data structures with only 34 distinct field names. This approach not only addresses specific challenges at Infologic, but also provides a valuable framework for organizations navigating the complexities of data lakes. We believe that concept lattices are effective visual tools that are accessible even to individuals who are not data analysis experts. They can read and understand the presented concepts and make informed decisions accordingly. Moving forward, the insights gained pave the way for a cleaner and a more exploitable data lake. Moreover, the resulting unified schema can serve as a model for our Electronic Data Interchange (EDI~\cite{DBLP:journals/jmis/PremkumarRN94}) module, a generic Copilote component serving as an interface for the exchange of data between Copilote instances and other systems. As a part of future work, a promising avenue is to incorporate advanced tools to enhance and automate our manual data analysis methods used in this paper. Integrating Natural Language Processing (NLP) and graph mining techniques will empower us to identify groups of similar fields, enabling the mapping of disparate fields to a unified name. Additionally, leveraging analogy-based reasoning will assist in identifying corresponding fields across different tables. For example, the following analogy would make it possible to unify the fields $a$ and $b$: ``the field $a$ is to the table $A$ as the field $b$ is to the table $B$''. Thanks to these techniques, our approach would be able to scale to larger data lakes with a higher number of data structures to consolidate. Another possibility to improve the scalability of our approach is to exploit AOC-posets~\cite{DBLP:conf/cla/DolquesBH13}, which are smaller and more concise subsets of concept lattices.

\section*{Acknowledgement}
This research was supported by the ANR project ``Formal Concept Analysis: A Smart Tool for Analyzing Complex Data'' (SmartFCA), ANR-21-CE23-0023. The authors also express their sincere gratitude to Philippe Cancellier and Guillaume Kheng for their contributions to the datalake at Infologic and to Amedeo Napoli for insightful discussions.
\bibliographystyle{splncs04}
\bibliography{biblio}

\end{document}